# Iceline variations driven by protoplanetary disc gaps

Madelyn Broome ,[1,2]★ Mihkel Kama ,[3,4] Richard Booth [5] and Oliver Shorttle[1,6]

[1]*Institute of Astronomy, University of Cambridge, Madingley Road, Cambridge, CB3 0HA, UK*
[2]*Department of Astronomy and Astrophysics, University of California Santa Cruz, 1156 High Street, Santa Cruz, CA 95064, USA*
[3]*Department of Physics and Astronomy, University College London, Gower Street, London WC1E 6BT, UK*
[4]*Tartu Observatory, University of Tartu, Observatooriumi 1, 61602 Tõravere, Tartumaa, Estonia*
[5]*Astrophysics Group, Imperial College London, Prince Consort Road, London SW7 2AZ, UK*
[6]*Department of Earth Sciences, University of Cambridge, Downing Street, Cambridge, CB2 3EQ, UK*

## ABSTRACT

The composition of forming planets is strongly affected by the protoplanetary disc's thermal structure. This thermal structure is predominantly set by dust radiative transfer and viscous (accretional) heating and can be impacted by gaps – regions of low dust and gas density that can occur when planets form. The effect of variations in dust surface density on disc temperature has been poorly understood to date. In this work, we use the radiative transfer code MCMax to model the 2D dust thermal structure with individual gaps corresponding to planets with masses of 0.1 $M_J$ –5 $M_J$ and orbital radii of 3, 5, and 10 au. Low dust opacity in the gap allows radiation to penetrate deeper and warm the mid-plane by up to 16 K, but only for gaps located in the region of the disc where stellar irradiation is the dominant source of heating. In viscously heated regions, the mid-plane of the gap is relatively cooler by up to 100 K. Outside of the gap, broad radial oscillations in heating and cooling are present due to disc flaring. These thermal features affect local dust–gas segregation of volatile elements ($H_2O$, $CH_4$, $CO_2$, and CO). We find that icelines experience dramatic shifts relative to gapless models: up to 6.5 au (or 71 per cent) closer to the star and 4.3 au (or 100 per cent) closer to the mid-plane. While quantitative predictions of iceline deviations will require more sophisticated models, which include transport, sublimation/condensation kinetics, and gas–dust thermal decoupling in the disc atmosphere, our results suggest that planet-induced iceline variations represent a potential feedback from the planet on to the composition of material it is accreting.

## 1 INTRODUCTION

The composition of a planet is determined by the relative proportion of solids and gas it accretes and by the elemental composition of those solid and gaseous reservoirs (Öberg, Murray-Clay & Bergin 2011; Williams & Cieza 2011). In the simplest view of planet formation, the elemental distribution between the gas and solids is determined by the planet's formation location. The import of formation location is two-fold: (1) the disc temperature at and around the planet's location determines the distribution of volatiles between gas and ice phases (e.g. Boss 1998; Lesniak & Desch 2011; Fedele et al. 2016) and (2) disc kinematics induced by, e.g. giant planets in the disc (Cridland, Bosman & van Dishoeck 2020), will influence availability and delivery of different volatile molecules. Here, we focus on the impact of disc temperature structure on the distribution of volatiles between gas and solid phase.

The local disc temperature is predominantly controlled by (1) radiative transfer and the stellar radiation field and (2) the viscous dissipation as material is accreted on to the star (Kenyon & Hartmann 1987; Chiang & Goldreich 1997; Lesniak & Desch 2011; Woitke 2015). Viscous heating tends to dominate the local energy budget at the mid-plane near the star – where the material is densest (Lesniak &

Desch 2011; Rafikov 2017). At lower densities − higher up and further out in the disc − the optical depth of the dust and gas is too low for viscous heating to have a significant impact on local temperature (Garaud & Lin 2007). Instead, the energy budget of the upper layer of the disc is dominated by stellar irradiation (for a review of disc models see, e.g. Dullemond et al. 2007).

This simplified layered model of disc heating provides a helpful framework for understanding the variance of the disc's thermal structure in the presence of surface density variations. Since both radiative and viscous heating are sensitive to variations in the local surface density and disc geometry (Dullemond et al. 2006; Garaud & Lin 2007; Piso et al. 2015) – and planets have been shown to induce such variations − the thermal structure of a protoplanetary disc (PPD) may therefore be perturbed by the presence of a planet. In this contribution, we use radiative transfer calculations in a planet-hosting disc to understand the temperature anomalies planets induce in their disc and the ensuing effect on iceline locations.

Massive enough planets cause decreases in the gas surface density local to their orbit (Lin & Papaloizou 1986). This dip in gas surface density drives a corresponding decrease in the dust surface density, but one that is orders of magnitude larger, potentially resulting in the well-documented rings and 'gaps' in millimetre continuum emission from discs (e.g. Crida, Morbidelli & Masset 2006; Duffell 2015; Dipierro & Laibe 2017; Zhang et al. 2018a). While not all gaps are necessarily the result of planet formation, planet formation is expected to almost always induce a gap in the disc's gas and dust (e.g. Malik et al. 2015; Van der Marel, Williams & Bruderer 2018).

Dust is of particular interest in discussions of radiative heating and cooling, because, despite only accounting for ∼ 1 per cent of the disc's mass, dust accounts for the majority of the radiative transfer in the disc (Chiang & Goldreich 1997). Thus, gaps in the gas and dust distributions are likely to have an effect on the temperature of the planet-forming environment.

A gas gap opens when the planet's local gravitational torque exceeds the viscous torque in the disc (e.g. Goldreich & Tremaine 1980; Crida et al. 2006; Dipierro & Laibe 2017). A combination of torque balancing, radial drift, and pressure trapping (Tanaka, Takeuchi & Ward 2002; Fouchet et al. 2007; Rosotti et al. 2016) results in a gap, which is depleted of both gas and dust − of all grain sizes (see, e.g. Takeuchi, Miyama & Lin 1996; Brauer, Dullemond & Henning 2008; Watson et al. 2009; Birnstiel, Klahr & Ercolano 2012; Marel et al. 2018).

Since the dust surface density affects the propagation of radiation through the disc (Chiang & Goldreich 1997; Jang-Condell & Sasselov 2004), we can qualitatively consider several effects of the gap opening on disc temperature structure:

(i) Lower dust density above the mid-plane could decrease photon scattering into the gap, resulting in a cooler region;
(ii) Shadowing of the disc by the gap walls could also cool the mid-plane (Siebenmorgen & Heymann 2012);
(iii) Conversely, the decreased opacity in the gap could allow radiation to penetrate deeper into the disc, warming the normally shielded mid-plane (Dullemond et al. 2020).

With the interplay of these multiple complex radiative effects, it is difficult to predict the thermal impact of a gap on the disc without modelling radiative transfer through the disc (Jang-Condell & Sasselov 2004). Similar work by Alarcón et al. (2020) suggests that ring substructure would, in fact, induce warmer gaps and lead to gaps becoming volatile-enhanced regions, but they did not induce viscous heating in their model. Our approach further differs in that we turn to a radiative, rather than thermochemical, model.

Despite the importance of thermal structure for icelines and planet formation, hydrodynamical models of PPDs with gaps frequently assume an isothermal disc or a disc whose temperature is defined by a simple radial power law relationship. This choice is motivated by the computational expense of modelling a changing temperature structure in a hydrodynamical code. The choice of a locally isothermal disc is also often justified by observations of outer discs (> 100 au), which are dominated by stellar irradiation (Ziampras, Kley & Dullemond 2020). However, Ziampras et al. (2020) found that, up to 100 au, hydrodynamical models of gap opening and morphology in a radiative (versus isothermal) disc produce gaps closer to observations than an isothermal disc does.

The more nuanced picture of thermal variations that radiative transfer models can provide is also necessary to accurately model icelines (Sasselov & Lecar 2000). The iceline marks the physical points in the disc at which the volatile transitions from gaseous to solid (ice) (Hayashi 1981). A change in the local or global temperature in the disc changes the segregation of volatiles between the dust and gas. Thermal structure changes can alter the path of the condensation front, or, snowline/iceline, of those elements such as $H_2O$, $CH_4$, $CO_2$, and CO through the disc (see, e.g. Garaud & Lin 2007; Harsono, Bruderer & van Dishoeck 2015; Piso et al. 2015; Panić & Min 2017).

The state of the volatile elements during planet formation is predicted to have a dramatic effect on the composition of a planet's core and atmosphere (e.g. Öberg et al. 2011; Bosman et al. 2021). In our own Solar system, the gas giants were likely formed beyond the $H_2O$ snowline, where water was present as ice (Boss 2000; Sasselov & Lecar 2000; Kruijer et al. 2017). These planets are starkly different in composition, mass, and radius to the terrestrial planets, which were formed interior to the $H_2O$ iceline.

In addition to influencing the ultimate composition, volatile abundance, and size of planets, the presence of volatile ices in dust has been suggested to play a role during the planet formation process (Brauer et al. 2008; Ida & Lin 2008; Min et al. 2011). As the gases condense on to dust grains, the grains gain a new icy mantle, which increases the grains' 'stickiness' and changes their coagulation properties (Apai & Lauretta 2010; Min et al. 2011; Lee 2019; Van der Marel 2019). Though theories of planet formation vary, all require an accretion or collision phase in which the planet collects solid materials (e.g. Goldreich & Ward 1973; Boss 1998). Volatile redistribution triggered by gap-opening and associated thermal structure changes in the disc may therefore have a feedback on the growth of planets.

Modelling variations to the paths of volatile icelines in the presence of the gap gives a good qualitative metric of the gap's relative impact on the local and global disc temperature structure.

In order to understand the effects of the mass and semimajor axis of a gap-forming planet on a disc's temperature structure, we model gaps corresponding to planets with masses from 30 $M_{\oplus}$ to 5 $M_J$. Each model contains a single gap and exhibits a number of consistent thermal features, such as intra-gap heating and cooling, thermal noise in the disc's inner mid-plane, and hot–cold features extending past the outer edge of the gap. We pay particular attention to the temperatures of the mid-plane in the gap, of the gap walls, of the regions adjacent to either side of the gap, and, broadly, the inner and outer disc. Overall, we find that gaps have a significant effect on the temperature structure of the disc both inside and outside of the gap.

Relevant details of the selected gap model are presented in Section 2.2. How MCMax models disc physics is detailed in Section 2.1 and the solutions it generates, including temperature maps and thermal features, in Section 3. Finally, the implications of the thermal structure for icelines and planet chemistry are explored in Section 4.

## 2 METHODS

### 2.1 Modelling radiative transfer in the disc

MCMax is a customizable 2D/3D Monte Carlo dust radiative transfer code that specializes in accurately calculating the vertical structure of PPD's (Min et al. 2009). It is possible to define the disc properties (mass, size, dust surface density, and grain compositions), stellar properties (mass, luminosity, and temperature), and computational parameters (number of grid cells, number of photon packets, and number of iterations towards equilibrium). The total luminosity ($L_{tot}$ = stellar + viscous heating) is divided into $N_{phot}$ packets each carrying luminosity equivalent to $L_{tot}/N_{phot}$[1] (Min et al. 2011).

In the 2D mode, MCMax assumes a disc that is axisymmetric and symmetric across the x-axis. A 2D slice of the upper half of the disc is then divided into grid cells that vary in size as a function of $r$ (the x-axis) and $\theta$ (the y-axis). MCMax sets the grid adaptively to minimize errors that arise due to low photon counts. Cell widths range from <1 au in the inner disc to 12 au in the outer disc, with average heights in $\theta \sim z/r$ space of 0.0125 for most of the disc's height. In the disc's upper atmosphere, above $0.6 < z/r < 0.8$, the grid cells can be as tall as $\delta(z/r) = 0.033$. These grid dimensions reflect the number of photons present in each of these regions. The code is not time evolving, but does iterate through dust settling (per Dubrulle, Morfill & Sterzik 1995) and vertical structure calculations until the disc reaches equilibrium for the given initial conditions (Fig. 1).

---

[1] MCMax uses the Bjorkman & Wood (2001) scheme, where the frequency of a photon package is modified in each absorption-and-re-emission event, but the total energy each package carries remains the same.

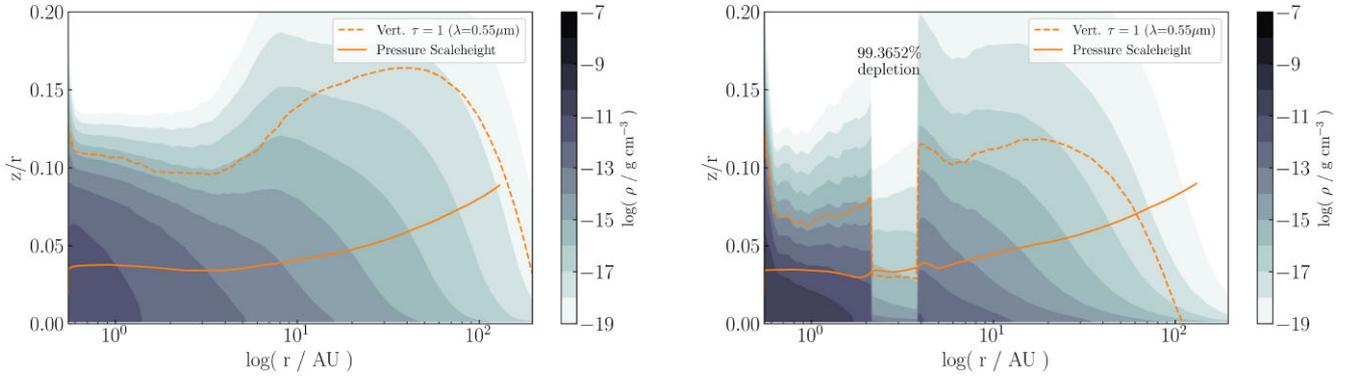

**Figure 1.** Dust mass density for a gap corresponding to a 1 $M_J$ planet at 3 au − fiducial gap-free model (left-hand panel) and gap model (right-hand panel). Vertical $\tau = 1$ surface for $\lambda = 0.55$ nm photons (peak of the solar spectrum) is plotted in orange dashed lines. Solid orange line is the pressure scale height computed by MCMax. Gas density is taken to be $100 \times \rho_{dust}$ everywhere.

MCMax does not include line cooling, nor does it model hydrodynamic heat transfer mechanisms like convection. The disc energy budget in MCMax is computed using the two dominant sources: viscous heating and radiative heating/cooling.Temperature contributions of the steady-state, time-independent accretion on the star are computed at each cell in the disc by assuming that the disc radiates as a blackbody (Min et al. 2009). Users have the ability to define the accretion rate ($\dot{M} = 10^{-8} M_\odot$, in our case) and total number of photon packets in the disc.

In well-shielded, optically thick regions like the inner disc midplane, the noise that emerged in the thermal maps (Fig. 2) led us to use $10^7$ photon packets to minimize errors in these regions, which have low photon counts. While some noise is still present for $10^7$ photons, it is greatly reduced from the $10^6$ photon case and can be reduced further by increasing the initial photon packet count to $10^8$ or, in the future, by employing the Lucy (1999) radiative transfer algorithm instead of Bjorkman & Wood (2001)'s. The computational expense of $10^8$ photon packets was too high for our purposes (runtimes on the order of 700 h for a single disc), but single iteration trials showed that the thermal features of the $10^8$ photon case are consistent with the $10^7$ case.

## 2.2 Implementing a gap

To understand the effects of varying disc density we model a disc of 200 au in radius, with a characteristic radius of 60 au, dust mass of $M_{tot} = 10^{-4} M_\odot$, and a density profile of

$$\rho_0(r) \propto \frac{\exp(-(z/H_s)^2/2)}{H_s\sqrt{2\pi}} \frac{M_{tot}}{2\pi(R_{out} - R_{in})} r^{-1},$$

where $z$ is the height above the mid-plane, $H_s$ is the scale height at the characteristic radius, $R_{in}$ and $R_{out}$ are the inner and outer radii of the disc: 0.55 and 200 au, respectively. We introduce a gap into this disc by modifying the initial dust surface density distribution according to the prescription of Duffell (2019). Gaps can be formed by a number of mechanisms (see, e.g. Marel et al. 2018) and the thermal features induced by a planet-generated gap may be representative of gaps of any origin. However, since icelines have a significant impact on planet formation and composition, we choose here to define gaps with dimensions corresponding to planets of given masses and located at particular semimajor axes and evaluate their thermal consequences in this context.

Duffell (2019)'s gap model is empirically derived from simulations of interactions between a non-migrating planet and a locally isothermal disc. The resulting analytic model gives the surface density of gas in a steady-state disc as a function of planet and disc properties. This gas surface density profile is translated to a dust profile using a dust-to-gas ratio of 0.01.

The resultant dust density gap varies in depth and width as a function of planet mass, semimajor axis, and radial location in the disc. Gap size also varies with mach number (inverse aspect ratio) and Shakura–Sunyaev $\alpha$-viscosity (Shakura & Sunyaev 1973). These final two parameters are eliminated as independent variables by selecting a constant $\alpha = 10^{-3}$ (Rafikov 2017) and computing a mach number, $\mathcal{M}$, which scales with the radius, $r$, of the fiducial disc.

Given our choices of independent parameters, the width of the gap depends solely on the semimajor axis, $a$, and the ratio of the planet to stellar mass, $q$. Here, the gap's width, $\Delta$, is defined as the radial distance over which the azimuthally averaged surface density is less than half of the initial surface density (Kanagawa et al. 2016). The resulting gap is asymmetric with $\Delta$ scaling as

$$\Delta/a \sim \sqrt{q\mathcal{M}}/\alpha^{1/8}, \qquad (1)$$

when $\Delta \ll a$. This empirically derived result varies by a small factor of $(\alpha/\mathcal{M})^{1/8}$ from similar analytic derivations by Kanagawa et al. (2015). Since $\mathcal{M}(r) = r/H(r)$, $\mathcal{M}(r)$ decreases as a power law with increasing radius, resulting in a non-linear relationship between the gap dimensions derived by Kanagawa et al. (2016) and those by Duffell (2019). However, simulations of gaps defined in both studies result in the same general thermal features in the discs.

In our investigation, we give the gap depth as a fractional depletion of dust in the gap at the location of the planet, $d(a) = 1 - \Sigma(a)/\Sigma_0$, where $\Sigma_0$ is the dust surface density at the edge of the gap. Thus, $d = 0.9$ refers to a gap that is depleted of 90 per cent of its dust. Details of the piecewise analytic function derived by Duffell can be found in Duffell (2019).

The gas gap model is predicted to be accurate up to masses of a few times Jupiter's mass. Above Jupiter's mass, $q \approx 10^{-3}$, the scaling of depth with planet mass also becomes much steeper (Ginzburg & Sari 2018; Duffell 2019). Duffell's prescription will break down when the gap becomes too deep or unstable. So, our range of planet masses are selected with these limitations in mind.

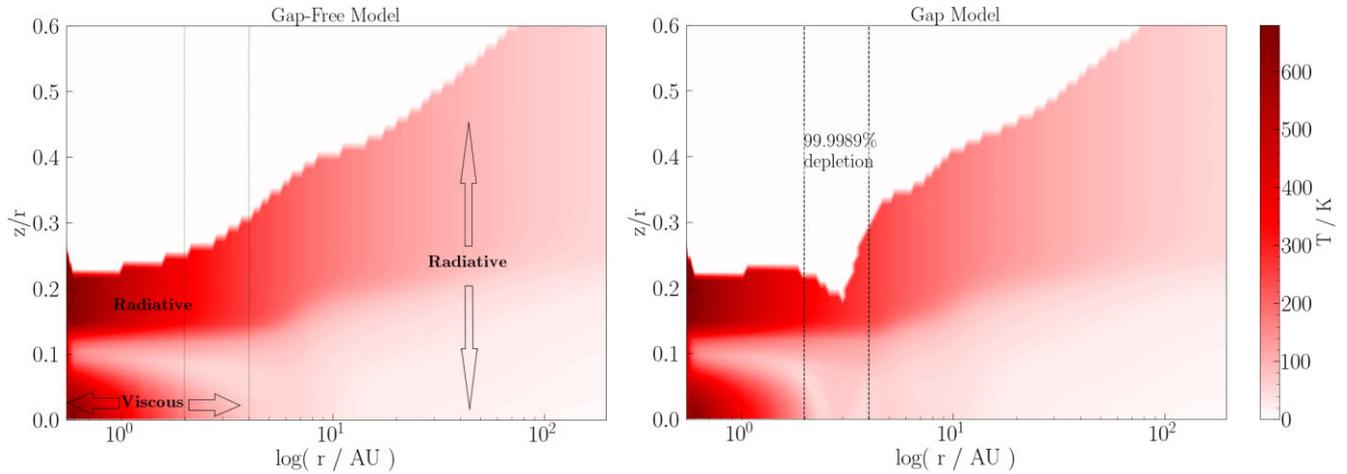

**Figure 2.** Raw disc temperature maps for fiducial Model (left-hand panel) and Jupiter-like gap model (right-hand panel) — the dust temperature in a 2D slice of the upper half of the disc (MCMax assumes axisymmetry and symmetry across the *xy*-plane) as a function of disc radius and $z/r$, the disc height normalized by radial location. The same colourscale is used for both maps. The width of the gap is plotted in dashed lines in the model with a gap (right-hand panel). The gap is depleted of more than 99 per cent of the dust at the location of the planet (3 au, 1 $M_J$). The same region is identified by light dashed lines in the gap-less fiducial model (left-hand panel). MCMax has a hardcoded density minimum of $10^{-50}$ g cm$^{-3}$, so all regions with $\rho_{dust} < 10^{-48}$ g cm$^{-3}$ are masked in white, resulting in an effective disc 'edge'. The hot mid-plane region in the lower left is due to accretion ($\dot{M} = 10^{-8} M_\odot$/yr) on to the star. The cooler horizontal slice $z/r \lesssim 0.1$ is radially optically thick ($\tau(r, z/r) > 1$).

The broad effects of relative planet mass and orbital radius on gap size can be summarized as follows: gap width and depth increase with increasing $q$, when all other parameters are held constant, and the gap width increases with $a$. In detail, the dust is depleted from the gap would be redistributed throughout the inner disc, accreted on to the planet, or trapped in the pressure bump that forms to the exterior of the gap. However, for simplicity, here we have assumed a constant dust-to-gas ratio throughout the disc. The effect of the pressure bump — in particular, the higher density and accumulation of larger grains in that region — is likely non-negligible on the heating and cooling outside of the gap (Alarcón et al. 2020), though, and merits future investigation.

The gaps modelled corresponded to a range of planet-star mass ratios from $q = 10^{-4}$ to $q = 5 \times 10^{-3}$ and of semimajor axes from 3 to 50 au. Since we find that the thermal effects of placing a gap are only significant in the inner 100 au, the results from our small model disc are also relevant predictors of gap-induced thermal variations in larger discs.

The modelled dust is composed of solid grains (Pollack et al. 1994; Min & Flynn 2010), each with a 20 per cent carbon makeup, 0.25 porosity, and with 10 sizes distributed from 0.1 to 3000 μm according to the MRN distribution (Mathis, Rumpl & Nordsieck 1977). We model a parametrized disc structure with DIANA standard opacities and polyaromatic hydrocarbons (PAHs) (Woitke et al. 2019). MCMax computes temperature dependent opacities for interactions with photons (0.05–5000 μm) – with a specific focus on non-destructive interactions with UV photons – and includes quantum heating.

In this work, we assume that all dust species are distributed homogeneously throughout the disc and are depleted in the gap in equal proportions. As previously mentioned, trapping of large dust grains in the pressure bump at the outer edge of the gap as well as dust settling mean that this is a simplifying assumption. Though MCMax does take into account dust settling, meaning that larger grains are concentrated towards the disc mid-plane, the code version used here assumes perfect thermal coupling between the gas and dust in all grid cells.

MCMax takes only stellar temperature, mass, and luminosity as independent variables. Since the code also does not include any high energy radiation beyond UV, we are unable to define a realistic young star – highly active with strong EUV and X-ray radiation. We, therefore, chose to model our star using the parameters of the current-day Sun.

It is worth noting that changing the input stellar properties from a G- to O-type star had minimal effects on the presence of the thermal features. However, without the ability to change the stellar spectra along with these properties, we are likely missing nuances that would be present in the thermal map of an O- versus that of a G-star. Nevertheless, the fact that the heating and cooling patterns persisted for a disc of constant size and composition across increased stellar temperature, mass, and bolometric luminosity suggests that the heating and cooling induced by a gap may be present in discs irrespective of stellar type – though such a conclusion requires further investigation.

## 3 RESULTS

Using MCMax, we generated 25 distinct models of a disc with a single gap of varying size and location and compared each with the model for gapless fiducial disc. The thermal features that emerge when the gap and fiducial models are differenced are well-described by a number of physical properties of the gap and disc, and can be understood as a consequence of the two-layer structure of viscous and radiative heating.

### 3.1 Inside of the gap

When the gap models and the fiducial gap-free model are differenced, almost all of the thermal features that emerge are confined to the radially optically thick region below the radial $\tau = 1$ surface for $\lambda_{\odot, max} = 0.5$ μm — the surface where most radially incident stellar photons interact. This surface is unaffected by the presence of the gap because the extent of the lower density dust across the width of the gap is insufficient to lower the radial column density enough to alter the path of the $\tau = 1$ surface. In contrast, the vertical $\tau = 1$ surface is below the radial $\tau = 1$ surface and is dramatically affected by the gap. The height of the vertical $\tau = 1$ surface drops to the mid-plane for most gap conditions we model. This vertical shift means that vertically scattered photons can penetrate much deeper into the disc than they can when no gap is present (Fig. 3). Several of the thermal features correlate well with the position of the vertical $\tau = 1$ surface.

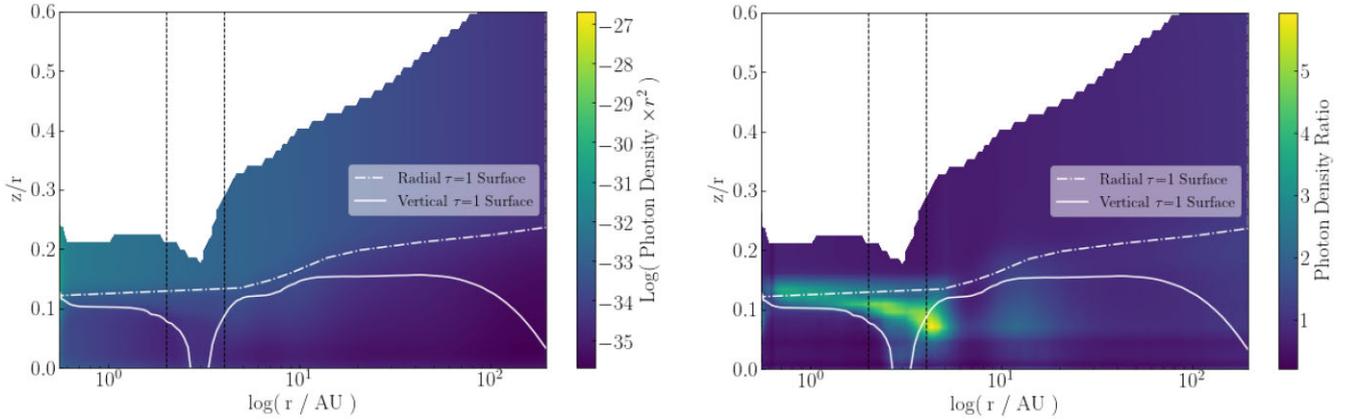

**Figure 3.** Photon density maps − for a 1 M$_J$ planet at 3 au. Left-hand panel: the log number of photons packets per unit volume is scaled by $r^2$ to give an approximate local photon flux ($F \propto L/r^2$) and allow for better visualization. Right-hand panel: the photon density ratio (number of packets in the gap model per unit volume relative to the fiducial model). Radial and vertical $\tau = 1$ surfaces are plotted in dashed and solid white, respectively. Vertical-dotted black lines indicate the width of the gap. Stripes of high photon density beyond the gap correspond to flaring-induced changes to the radial $\tau = 1$ surface and, therefore, to the thermal oscillations seen in Fig. 4. Dark horizontal stripes are artefacts of MCMax gridding.

Where the gap intersects with the inner disc ($0.55 < r < 4$ au, $z/r \lesssim 0.1$) where viscous heating dominates in the fiducial model, the gap has a net cooling effect. Fig. 2 demonstrates how the radial extent of viscous heating is truncated by the low density of dust in the gap. Viscous dissipation provides a set amount of energy per unit mass, so heat production per kg of dust is unaffected by the presence of a gap. However, the lower optical depth in the gap allows this energy to more easily escape and increases the cooling efficiency of the region resulting in a mid-plane that is relatively cooler than that in the gap-free fiducial model. Vertically scattered photons are still acting to warm this layer where it is optically thin (Fig. 4, the disc is warmer than the fiducial case above vertical $\tau = 1$ than below it); however, the heat added by these scattered photons is less than the heat lost by photons escaping from the mid-plane layer where viscous dissipation dominates. The observed net cooling inside of the gap shows how important viscous heating is to the mid-plane energy budget at in the inner disc. Though the gaps at 1 au were already well within the water condensation front for the discs we modelled, for warmer discs − where the water snowline would intersect the mid-plane at radii >1 au − it is possible that that net cooling of the viscous layer by gaps at low radii could lead to water condensing closer to the star than predicted by uniformitarian disc models. In such a case, properly modelling the gaps would be essential for predicting planet composition, since the availability of water ice to a growing planet could yield a wetter planet than a gapless thermal model would predict.

In the rest of the disc, at larger radii ($\gtrsim 4$ au) and higher above the mid-plane (in the disc's atmosphere, $z/r \gtrsim 0.1$), stellar irradiation dominates the local energy budget. The effects of stellar irradiation are difficult to predict without modelling, because lower density not only increases cooling efficiency, it allows for increased photon penetration deep into the disc. Just as photons may more easily penetrate into the disc, as can be seen in Fig. 3. Just as photons may more easily penetrate into the disc, they can also more easily escape. On balance, however, the heating dominates down to the mid-plane in gaps further out in the disc (Fig. 5c and d) and in the atmospheres of gaps found in the inner few au. (Fig. 5a and b).

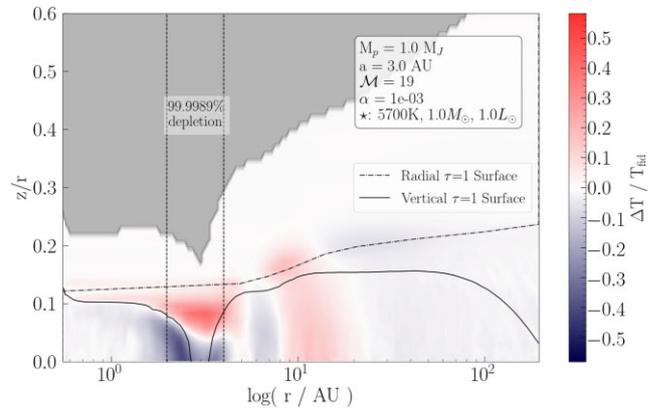

**Figure 4.** Scaled temperature residuals − aligning and differencing the temperature maps in Fig. 2 reveals the thermal effect of gap opening. Red regions are relatively hotter than the gapless fiducial model and blue are relatively cooler. $\Delta$T is the gap model minus fiducial temperature. The radial hot–cold oscillations inwards of the gap below $z/r = 0.1$ are noise due to low photon count in the well-shielded mid-plane. Vertical dashed lines indicate the edges of the gap (see, equation 1). The solid grey mask indicates where dust density $<10^{-48}$g cm$^{-3}$ and MCMax's radiative solution is considered invalid. Most thermal features occur in the radially opaque region below the radial $\tau = 1$ surface. Within the gap, the temperature-dependent vertical dust opacity (solid line) is much lower, allowing photons to penetrate into the disc, heating the material within. The relative cooling near mid-plane in the gap occurs only when the gap intersects with the region of viscous accretion from $0.55 < r < 4$ au. Low dust density in the gap increases cooling efficiency. We find that low density decreases vertical opacity, allowing greater photon penetration (Fig. 3), which does lead to some warming: dust within the gap and below the vertical $\tau = 1$ surface is colder than dust above it (see ∼$z/r = 0.05$ and $r = 3$ au). The sharp transitions between heating and cooling regions agree with the transitions between regions of very high and low photon density (Fig. 3). Note that we do not see the same intra-gap upper atmosphere heating that Alarcón et al. (2020) find.

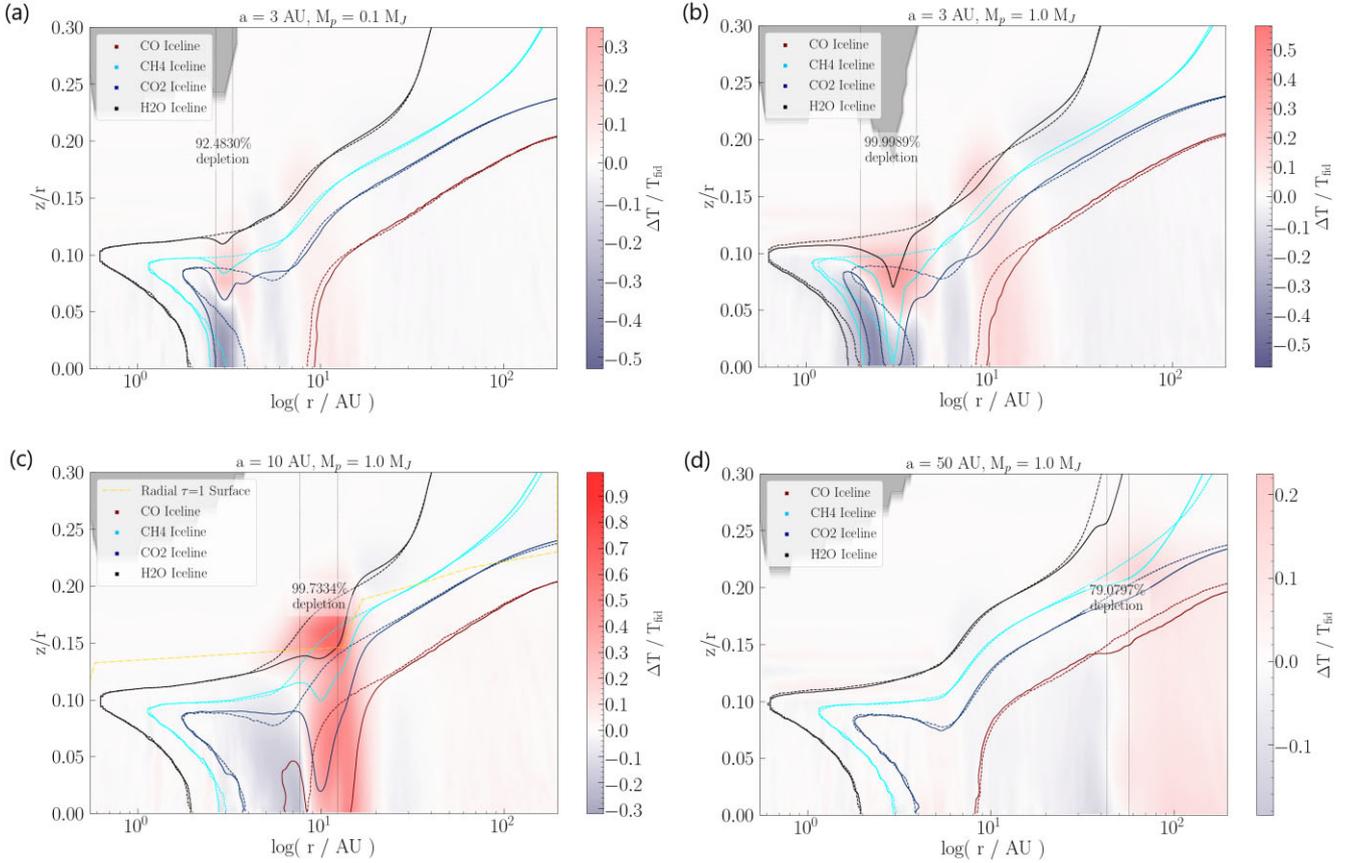

**Figure 5.** Iceline variations in the presence of a gap – icelines of $H_2O$ (black), $CH_4$ (cyan), $CO_2$ (navy), and CO (dark red) for identical discs with a gap (solid line) and without (dashed) are plotted as the locations in the disc where the disc local temperature equals the sublimation temperature corresponding to the density of each volatile at that location (assuming Solar system volatile abundances). The gaps shown correspond to a 0.1-$M_J$ planet at 3 au (a) and a 1-$M_J$ planet at 3 au (b), 10 au (c), and 50 au (d). The radial $\tau = 1$ surface is given as a yellow dash–dotted line in (c). Relative temperature differences are plotted as in Fig. 4 with the same colourscale for all discs. Grey regions are those where dust density $< 10^{-48}$ g cm$^{-3}$ and MCMax's radiative solution is considered invalid. The magnitude of the dips (downward vertical translations) in the iceline paths increase with the depth and width of the gap (a) and (b). Even in cases the gap does not intersect the iceline, it can experience horizontal shifts on the order of an AU (e.g. b). Retreat behaviour (shifting to higher radii) can be seen in the case of the CO iceline. Advancement behaviour (shifting to lower radii) in, e.g. the $CH_4$ snowline is significant for (a) and (b).

Regardless of where the gap is located, dust in regions which are optically thin to vertically scattered photons is warmer than dust in vertically optically thick regions, even when cooling dominates (Fig. 4), revealing that the net effect of increased photon penetration is heating. For distant gaps, this heating can reach all the way down to the mid-plane, resulting in temperature increases of up to 50 per cent compared to the gap-free cases or tens of degrees Kelvin at the mid-plane (Fig. 5c and d). Since line cooling is not included in MCMax, the greater number of incident photons into the low density dust of the gap is the most probable cause of this heating. Photon maps in Fig. 3 reveal that there is indeed a greater number of photons per unit volume in the gap than within the same region in the fiducial model.

The magnitude of relative heating in the gap increases with increasing gap width and depth (Fig. A1). Additionally, when we plot the location of the gas that experiences the highest relative temperature increase (defined as $\Delta T/T_{fiducial}$) as a function of semimajor axis, we see that hotter material is found higher and further towards the outside edge of the gap the further the gap is away from the star. This is likely the result of greater disc flaring with increasing radius, which exposes material in the gap's upper atmosphere to more radially travelling stellar photons. Indeed, in the 10 au and 50 au gaps, the region of greatest heating traces the radial $\tau = 1$ surface almost exactly (Fig. 5c). Thus, the hottest region in the gap is that, which is *both* vertically and radially transparent, confirming that photons are a significant source of the intra-gap heating.

Even though the hottest region is always located well above the mid-plane and towards the outer edge of the gap, a planet at the centre of one of these gaps would still feel thermal effects of the gap. In Fig. 6, the relative temperature at the location of the planet, $T(R_P)/T_{fid}(R_P)$, is correlated with the location of the gap, as well as with the size of the planet that may have formed it. All gaps experience a relative temperature change at the planet's location, increasing with increasing planet size, except for when the very largest planet sizes (2.6 and 5.2 $M_J$) are reached. Larger planets imply a deeper gap. So, as a function of gap depth (Fig. 6b), the temperature at the mid-plane also increases relative to that of the fiducial model. However, once the gap has $< 0.001$ percent of the original mass in dust remaining, $T(R_P)/T_{fid}(R_P)$ decreases, suggesting that, at these extremely low densities, the cooling efficiency is finally high enough to have a significant effect on the net temperature of the gap. This behaviour accounts for the decrease in temperature with the very largest planet masses.

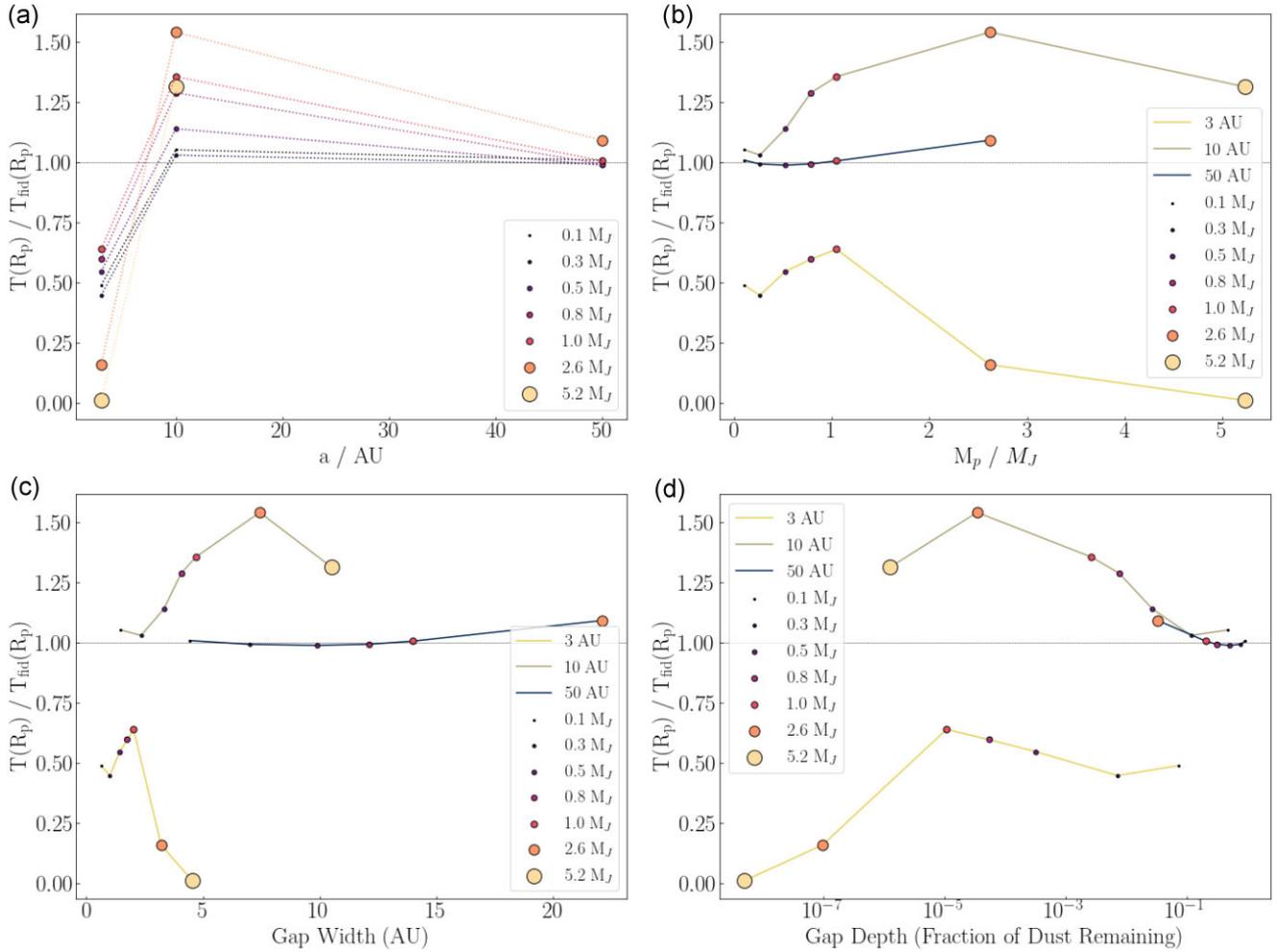

**Figure 6.** Relative thermal variations at the planet location – ratios of the temperature at the disc mid-plane ($z/r = 0$) at $R_P$ plotted as function of gap radial location (upper left-hand panel), planet mass (upper right-hand panel), gap width (lower left-hand panel), and gap depth (lower right-hand panel, where the depth increases to the left). Gap depth and width encode mass and radial location per the relations in Section 2.2. Increasing planet mass is given as increasing point size and is mapped to warm colours from dark maroon to yellow (also plotted as dashed lines in the upper left-hand panel). The gap radial location is indicated by solid lines colourized from yellow (3 au, larger incident stellar flux) to navy (50 au, smaller flux). The planet location is relatively warmer due to the presence of the gap when $T(R_P) > T_{fid}(R_P)$ (black horizontal line). The 3 au gaps (yellow lines) are relatively cooler due to the disruption of the warming mid-plane accretion layer. Otherwise, the mid-plane generally grows warmer as a function of larger gaps, though there is a downward trend in relative heating for the largest planets (a.k.a., deepest and widest gaps) that is independent of gap location. In these cases, $< 0.001$ per cent of the original mass in dust remains, a low enough density that cooling efficiency begins to be more significant.

The behaviour of $T(R_P)/T_{fid}(R_P)$ with changing radius is not as clear cut. For gaps far out in the disc (10 and 50 au), the further out the gap is, the smaller the temperature increase in the gap is. However, we find that the mid-plane of gaps in closer to the star are relatively cooler (Fig. 4) than it would have been in the gapless case. So, the $T(R_P)/T_{fid}(R_P)$ does not decrease monotonically with increasing $R_P$ (Fig. 6a).

### 3.2 Outside of the gap

The presence of a gap influences the temperature of the dust and gas outside of the gap as well. Two major features emerge when we examine the thermal maps.

The first is mild mid-plane noise present for both gaps in the inner disc (where both viscous and radiative heating are significant) and those further out in the disc (where the disc is purely radiative) (e.g. Fig. 4). The dust near the mid-plane between the star and the gap is consistently relatively cooler, but is also consistently noisy. This noise is the result of poor photon statistics and high temperature errors and occurs in the dense, well-shielded inner few au regardless of the location of the gap or the presence of mid-plane accretion.

The second outside-of-gap thermal variation is only present for gaps, which are further out in the disc, where heating is dominated by stellar radiation. The gap creates radial temperature fluctuations at larger radii in the disc. Moving radially outwards from the gap, heating in the gap transitions to a thin vertical stripe of relatively cooler dust and then back into a larger patch of relatively warmer dust (Fig. 4). The size of the cool stripe decreases with increasing gap width. Furthermore, when the gap is sufficiently far out in the disc (e.g. for the 50 au trials), the cool stripe disappears from the outer gap and, instead, we see the thermal oscillations occurring in the inner disc (Fig. 5).

These oscillations have similar origins as thermal instability/density waves in the disc (e.g. Ward 1986; Watanabe & Lin 2008;

Wu & Lithwick [2021]). The narrow band of material outside of the outer wall of the gap is less flared, which means that less radiation is incident on that region of the disc. The result is a thin, relatively cooler stripe of dust which can be seen in Fig. 4 at 7 au. This collapsed region of the disc then allows the outer adjacent region to intersect more photons, causing that region to heat up and flare. This might account for the relatively hotter region to right of the cool stripe in Fig. 4. The gradient of the radial $\tau = 1$, as well as the visible oscillations in the flaring of the disc 'edge' around 10 au in Figs 2 and 3 seem to support this interpretation.

The presence of thermal variations outside of the gap demonstrates the gap's influence on the temperature structure not just within gap, but in the PPD as a whole. This wide-reaching thermal influence has implications for icelines in the presence of a gap, which we discuss below.

### 3.3 Icelines

The heating both within and outside of gaps causes significant deviations to the locations of icelines in the disc. Fig. 5 shows the icelines for four volatile elements ($H_2O$, $CH_4$, CO, and $CO_2$). Below the iceline for a given volatile, most of the material is expected have 'frozen out', or sublimated from its gaseous to its solid state. To compute the pressure-dependent sublimation temperatures, we used equation (4) of Hollenbach et al. ([2009]), which gives the sublimation temperature due to thermal desorption physics. For these calculations, we used the adsorption/binding energies collated in Hollenbach et al. ([2009]) and Solar system abundances of the volatiles (Acke & van den Ancker [2004]; Ueda et al. [2005]; Marboeuf et al. [2014]). To give an approximate idea of the values, the pressure-independent sublimation temperatures given by (Lodders [2003]) are as follows: $H_2O$ (121 K, without graphite condensation), CO (21 K), $CH_4$ (41 K), and $CO_2$ (50 K).

Due to the presence of the gap, the icelines experience vertical and horizontal shifts in position relative to the path of an iceline in the fiducial gapless disc. These shifts can be classified into two main types: a dip and a retreat/advance. An iceline dips down when its path crosses the gap, resulting in a vertical downward translation relative to the fiducial iceline. The paths can dip from their fiducial path anywhere from 10 per cent to 100 per cent of the distance to the mid-plane. The result is deviations of up to 1 au vertically, allowing gas phase volatiles to be present much closer to the mid-plane than a gapless model would predict.

As the gap size increases, the dip becomes more pronounced (Fig. 5). This dipping is a direct consequence of intra-gap heating, as the magnitude of the relative heating in the gap also increases with increasing gap depth and width (Appendix A1).

The iceline deviations are not limited to the gap location, however, as even the CO iceline, which does not intersect with the gap centered on 3 au still feels the effects of the gap. In contrast to the $H_2O$ line, the CO iceline advances far from the star and near the mid-plane of the disc, where the temperature is much lower.

We term the horizontal shift of these icelines to higher radii (further from the star) 'retreat' and, to lower radii (closer to the star), 'advance'. In these cases, the iceline retreats from the fiducial path by a patch of relative heating (e.g. the thermal oscillations that occur adjacent to the gap's outer edge in Fig. 4), or, it advances to lower radii by tracing a patch of cooler material (such as to the left of the gap in Fig. 5).

The retreat behaviour is the more dramatic of the two. As gaps are placed at increasing semimajor axes and approaching an iceline of interest, the heating seen in Fig. 4 at ∼ 10 au drives that iceline to a higher radius.

The advance to smaller radii occurs when the snowline is close to one of the cooler stripes of material (e.g. Fig. 4; 6–8 au) and can trace a path along the lower temperature material. Icelines can advance up to 4.5 au closer to the star in these cases. Towards the interior of the disc, the $H_2O$ iceline advances up to 0.8 au towards the inner edge of the disc. Just as the magnitude of the dip increases with increasing gap size, the advances and retreats increase when the gap is wider and deeper.

## 4 DISCUSSION

We have found that the thermal effects of the gap are not limited to the gap's interior. The temperature of dust outside of the gap is also influenced by the presence of a gap. The alternating cooling and heating stripes adjacent to the gap's outer edge can result in horizontal translations of a few AU of the $CH_4$ and CO snowlines. This implies that planets forming at higher radii will feel the thermal effects of a planet forming at a lower radius and vice versa. Since not all gaps may be the result of planet formation, the fact that the rest of the disc experiences thermal variations indicates broadly the importance of even non-planet-forming gaps on planet formation.

### 4.1 Model limitations

Introducing more gaps per disc to correspond to the presence of additional planets will doubtless complicate the thermal features exhibited in these models. However, we have already seen that a gap in the inner disc may even affect, via snowline retreat, the material available to planets forming further out in the disc. The maximal magnitude (4.5 au) of the shifts seen was equivalent to the present-day average separation between the orbits of Jupiter and Saturn (4.3 au).

Since there is still one potentially prominent source of intra-gap mid-plane heating – accretion by the planet itself – missing from our models, it is yet unknown how accurate the radii of the icelines at the disc mid-plane in Fig. 5 are. Furthermore, given the limitations of MCMax, the model detailed is based on the current-day, low-activity sun, which means that our models are missing any potential heating due to extreme XUV stellar radiation at early times (see, e.g. Ayres [1997]; Sanz-Forcada et al. [2011]).

Though we are unable to simulate the X-ray photons that a hot young star would emit, simply increasing stellar temperature and luminosity produces discs with qualitatively the same heating and cooling patterns. A hotter star would, however, have very different iceline paths to those in discs around cooler stars. In O, B, and A stars, volatile icelines may occur too far out in the disc to either cross or be significantly affected by gaps located in the inner few au. While Agúndez et al. ([2018]) find that T Tauri *inner* discs have higher abundances of volatile molecules than those around hotter stars, in the *outer* disc their abundances are similar. These models suggest the difference in volatile abundances between these two pre-main sequence star types is minimal for our purposes. The higher UV flux and warmer discs of Herbig stars imply, however, that the icelines lie further out in the disc than for T Tauri stars (Agúndez et al. [2018]). This is confirmed by observations of Qi et al. ([2015]) and Agúndez et al. ([2018]), and implies that there may be an observable difference in the icelines/thermal structures of Herbig and T Tauri PPDs. The warmer overall disc of a Herbig star suggests that the relative cooling that results when a gap intersects with the viscous layer will likely be unable to cool the inner disc enough to extend the icelines further in towards the star. A T Tauri disc, on the other hand, would likely experience the full suit iceline deviations discussed in Section 3, which occur as a result of both the gap and icelines intersecting with the viscous mid-plane layer close to the star.

So, while the thermal features we see may be broadly the same for a hot young star, the magnitude and locations of those thermal changes should be different for a true pre-main sequence star. Therefore, rather than robust predictions of the exact segregation of volatiles between the dust and gas, these icelines should be considered as indicators of the broad thermal effects of a gap in a radiative disc.

### 4.2 Thermal variations induced by a Jupiter-like planet

Fig. 5(b) suggests that, if Jupiter was formed around 3 au as some theories posit (e.g. Lin & Papaloizou 1986; Desch, Kalyaan & O'D. Alexander 2018), the icelines in its vicinity would see significant deviations from the paths predicted by gapless models.

As it crosses the gap, the water snowline dips 37.5 per cent, or 1.1 au, down towards the mid-plane. In our model, this does not change the fraction of water ice available at mid-plane. Nor does the advance of the snowline by 0.25 au. However, Cridland et al. (2020) have suggested that meridonal transport and vertical accretion of material within one gas scale height of the planet may be significant in determining the solid volatile inventory available to planets. The water iceline does not fall within one gas scale height ($z/r = 0.025$), but the snowline's dip does increase the likelihood that water gas can be vertically transported down to the forming planet (Cridland et al. 2020).

If meridonal flows are significant, then perhaps the young Jupiter-like planet would be able to accrete some gas-phase $CH_4$, since the $CH_4$ and $CO_2$ icelines stretch 95 per cent and 100 per cent of the way down to the mid-plane within the gap, respectively. However, because the icelines also experience significant horizontal advances, if the planet were, instead, to accrete most of its material from the mid-plane, the planet would only have access to $CH_4$ ice, not gas. This result differs from predictions by the gapless model and has implication for planet formation. The segregation of the volatiles is important not only for the composition of the planet's atmosphere, but for the formation of the core (Stevenson 1982), with icy mantles on dust grains increasing their 'stickiness' and making coagulation more efficient (Chokshi, Tielens & Hollenbach 1993).

We see a similarly significant change in the $CO_2$ iceline when the gap is introduced. In the case of a 1-$M_J$ planet at 3 au, a gap-free model radiative model would predict that the planet formed outside of the condensation front and is therefore accreting mostly gaseous $CO_2$. When the gap is introduced the dip in the iceline as it crosses the gap means that gaseous $CO_2$ is still available at the planet's location at 3 au; however, the advancement of the iceline from 3.9 to 32 au puts ice-phase $CO_2$ much closer to the planet's formation radius at 3 au (Fig. 5b). It also means that a reservoir of solid $CO_2$ is present near the gap's inner edge and likewise falls within 0.3 au of the accreting planet. The reservoir itself is interesting in that it places $CO_2$ condensation almost a full 2 au closer to the star than a gapless model would predict.

In the case of our own Solar system, the presence of the gap complicates the picture of what materials may have been available to a forming Jupiter. Furthermore, Jupiter's gap would likely have implications for other planets forming in the Solar system. As can be seen by the retreat of the CO iceline by 1.2 au around $r = 10$ au, a Jupiter-like planet at 3 au would also influence the volatiles available to the planets forming further out in the disc.

### 4.3 Observability of planet-induced thermal structures

Since the temperatures of the dust and gas in a disc's upper atmosphere have been shown to be decoupled (Kamp & Dullemond 2004; Facchini 2018), our assumption that the two are in thermal equilibrium breaks down. Accurate thermochemistry in the disc upper atmosphere is important to predict the observability of features; however, the significant deviation of icelines, both vertically and horizontally near the gap, may still have observational implications. Furthermore, the impact of dust temperature on the iceline is exponential, while the gas temperature only affects the ice–vapour balance as $T_{gas}^{1/2}$. The temperature differences, thus affect the availability of volatiles in the gas- or ice-phase to planets or planetesimals and may affect molecular line emission. Notably, for example, the CO present in the gap for a 1-$M_J$ planet at 10 au would not be ice-phase CO as predicted by the thermal structure of a disc without a gap, but would, instead, be gaseous CO.

The gas-phase C/O ratio in a sample of discs has been shown to correlate with the presence, or lack of, dust rings (Van der Marel et al. 2021). While that study did not address additional heating facilitated by the low opacity in dust gaps, there must be a sub-population of discs where gaps are close to 'full-disc equivalent' iceline locations, and in which the additional heating determines which ices are evaporated. As a specific example, high spatial resolution spectroscopy within the ALMA large programme MAPS (Öberg et al. 2021) identified radial variations of the C/H, O/H, and C/O ratios in the HD 163296 PPD on ∼100 au scales. Multiple Jupiter-mass protoplanet candidates associated with gaps have been identified within this radial range (Liu et al. 2018; Zhang et al. 2018b) and may be contributing to these variations in gas-phase composition via iceline perturbations. We note that existing observations may have had limited ability to detect the types of effects studied in this work. It is possible that localized impacts of individual planet gaps on the gas temperature and composition in HD 163296 have not yet been resolved, especially in the inner few tens of astronomical units where such impacts are expected to be the most significant.

## 5 CONCLUSIONS

The presence of a gap in a radiative PPD has non-negligible impacts on the thermal structure of the entire disc and, significantly, on the segregation of volatiles near a growing planet.

We modelled 25 radiative, Solar system-like dust discs each with a single gap of varying width a depth. The gaps modelled corresponded to planets of masses ranging from 0.1 $M_J$ (31 $M_\oplus$) to 5.2 $M_J$ and semimajor axes of 3, 5, or 10 au using the analytic relations for gas gaps defined by Duffell (2019) (which is then translated to a dust gap using a fixed dust-to-gas ratio of 1/100).

The 2D radiative transfer code MCMax was then used to compare the temperature and photon density maps for the 22 trials with those for a fiducial gap-less disc.

The main sources of heating in PPDs are viscous (accretional) and radiative. In this model, the temperature change due to accretion on to the star is computed everywhere in the disc. For our assumed accretion rate of $\dot{M} = 10^{-8} M_\odot$/yr, viscous heating dominates near the mid-plane ($z/r \lesssim 0.1$) for the inner 4 au of the disc. Radiative transfer dominates the energy balance in the rest of the disc. The low density of the dust in a gap yields a lower optical depth (specifically, in the vertical direction) and, therefore, allows photons to penetrate deeper into these optically thin regions and heat the remaining dust; however, a lower optical depth also allows photons to more easily escape, thereby cooling the dust. In our models, the vertical $\tau = 1$ surface is 80 per cent –100 per cent further down into the disc than in the gapless fiducial model.

The resulting thermal structure indicates that the net effect of a gap is to warm the dust throughout the gap, so long as the gap does not fall within the inner few au, where viscous heating is important. In that inner disc ($\lesssim 4$ au), the material in the gap exhibits a two-layered structure, with a heated atmosphere ($z/r > 0.05$), but a relatively cooler mid-plane. This cooler mid-plane feature likely emerges because of the increased cooling efficiency of the low optical-depth gap region: where heating of the disc occurs directly by viscous dissipation, that energy is transported away more efficiently when the optical depth is low due to gap opening, resulting in a mid-plane which is cooler than the gap-free model by up to 97 K.

In the gap atmosphere, the dominant heat source is not viscous, but stellar irradiation and the increased photon penetration results in relatively hotter material. The presence of warming both inside and outside of the gap correlates strongly with increased photon density and with optically thin regions of the disc. This net warming of the gap agrees with the thermochemical models of Alarcón et al. (2020), though our 2D radiative transfer models also find the cooler mid-plane in the inner disc where viscous dissipation dominates the mid-plane energy budget. Therefore, whether the gap's lower density results in relative heating or cooling depends on which heating mechanism was dominant at that location in the fiducial gapless model.

How much the gap is heated is a function of how large it is and where it occurs in the disc. The magnitude of the relative warming increases with increasing planet size, gap depth, and gap width. We find that this trend holds up until we reach gap depths where < 0.001 per cent of the mass in dust remains. At those extremely low densities, cooling efficiency becomes sufficient to begin counteracting heating due to increased photon penetration.

The heating and cooling effects are not confined to within the gap. A mechanism akin to thermal instability waves is likely responsible for the heating and cooling regions outside of the gap. When combined with the intra-gap heating and cooling, these thermal variations cause the paths of icelines – computed using desorption physics – to dip vertically up to 4.3 au or shift inwards and outwards horizontally up to 6.5 au. Thus, the presence of a gap, whether it is planet-forming or not, can alter the availability of volatiles to planets forming elsewhere in the disc.

## ACKNOWLEDGEMENTS

We thank Michel Min for the use of his MCMax code and our referee for the thoughtful comments.

## DATA AVAILABILITY

The MCMax code is not publicly available and is not currently supported. Requests for the data underlying this article may be directed to the corresponding author.

## APPENDIX A:

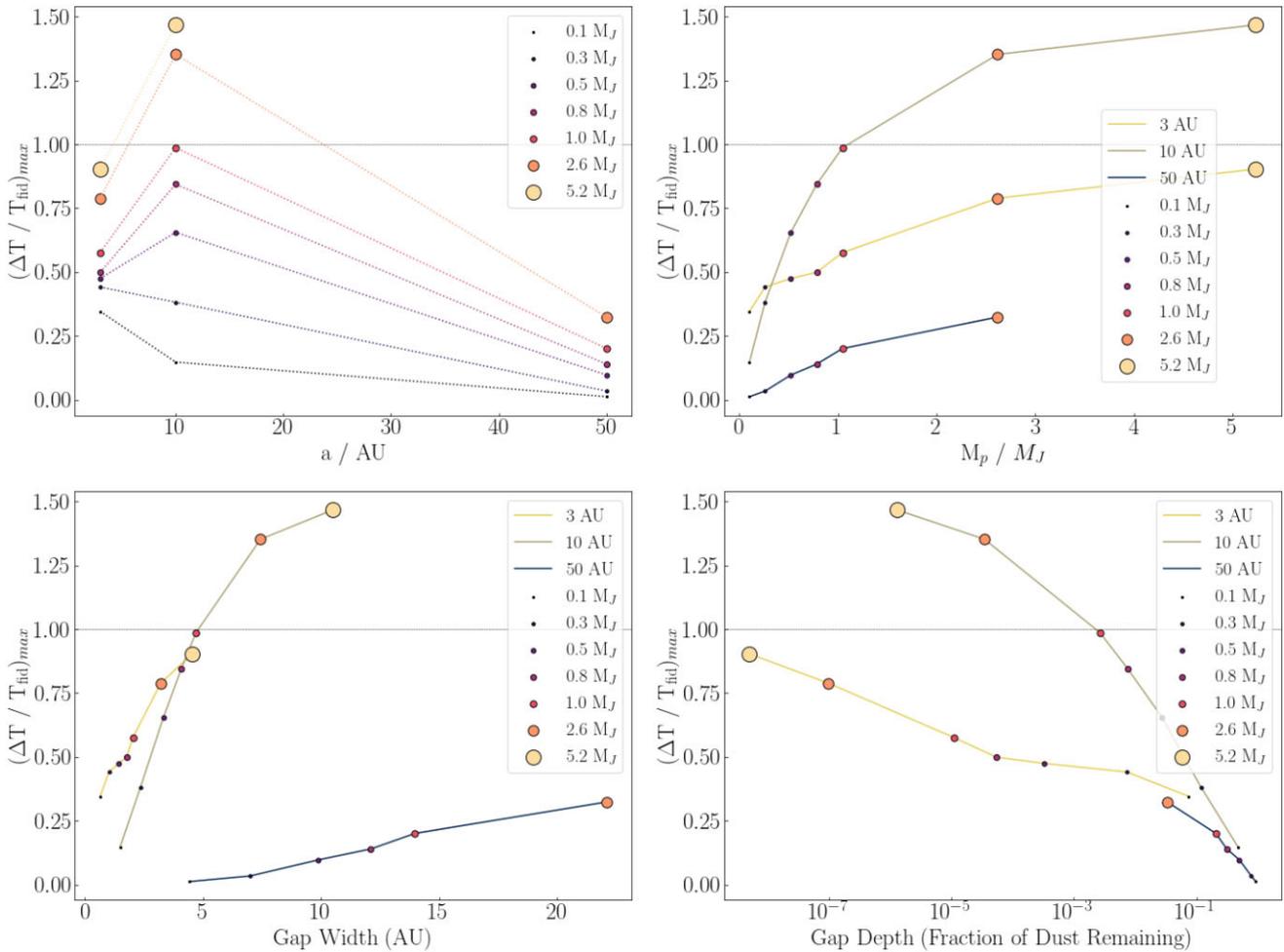

**Figure A1.** Maximum relative heating in gap − given as a function of planet semimajor axis (upper left-hand panel, colourized by planet mass), planet mass (upper right-hand panel, colourised by planet semimajor axis), gap width (lower left-hand panel), and gap depth (lower right-hand panel). Gap width and depth are a function of planet mass and orbital radius as described in Section 2.2. Colourization by mass reveals no obvious trends. Interestingly, $(\Delta T/T_{\text{bkl}})_{max}$ increases with increasing planet mass, but the middling gap at 10 au experiences a steeper and larger overall increase than the gap at 3 au (upper right-hand panel). This agrees with the large spread of heating magnitudes we see at 10 au as a function of planet mass (upper left-hand panel). At higher radii, the effect of a planet's mass on the heating diminishes. This large spread in heating magnitudes at 10 au appears to be a consequence of disc structure and not, as suggested by the trends in lower left- and right-hand panels, a function of the gap structure. The tighter distribution of planet temperatures at 3 au may be the result of the cooling mechanisms that come into play when the gap intersects the mid-plane accretion layer (Section 3.1).

This paper has been typeset from a TEX/LATEX file prepared by the author.